\documentclass[12pt, prd, showpacs]{revtex4}
\usepackage{amssymb}
\usepackage{amsmath}
\usepackage{appendix}
\usepackage{array}

\input{tcilatex}

\begin{document}

\title{Axially symmetric rotating \ black hole with regular horizons}
\author{H. V. Ovcharenko,}
\affiliation{Department of Physics, V.N.Karazin Kharkov National University, 61022
Kharkov, Ukraine}
\email{gregor_ovcharenko@outlook.com}
\affiliation{Faculty of Mathematics and Physics, Charles University, Ke Karlovu 3, 121 16
Praha 2, Czech Republic}
\author{O. B. Zaslavskii}
\affiliation{Department of Physics and Technology, Kharkov V.N. Karazin National
University, 4 Svoboda Square, Kharkov 61022, Ukraine}
\email{zaslav@ukr.net}

\begin{abstract}
We consider the metric of an axially symmetric rotating black hole. We do
not specify the concrete form of a metric and rely on its behavior near the
horizon only. Typically, it is characterized (in the coordinates that
generalize the Boyer-Lindquist ones) by two integers $p$ and $q$ that enter
asymptotic expansions of the time and radial metric coefficients in the main
approximation. For given $p,$ $q$ we find a general form for which the
metric is regular, and how the expansions of the metric coefficients look
like. We compare two types of requirement: (i) boundedness of curvature
invariants, (ii) boundedness of separate components of the curvature tensor
in a free falling frame. Analysis is done for nonextremal, extremal and
ultraextremal horizons separately. \newline
\end{abstract}

\maketitle

\section{Introduction}

The main feature of a black hole consists in the existence of a horizon. By
definition, it is implied that geometry is regular in its vicinity. Some
general conditions on the static metrics, even without requirement of
spherical symmetry, were formulated in \cite{v1} and extended to stationary
rotating axially symmetric space-times in \cite{v2}. The results of \cite{v2}
were generalized in \cite{tz}. Detailed classification of regular horizons
was suggested in \cite{prd08} for spherically symmetric metrics.

In both aforementioned works the properties of the metric were formulated in
terms of the proper distance. Meanwhile, much more usual and convenient
coordinate systems for rotating axially symmetric space-times represent
natural generalization of the Boyer-Lindquist coordinates which were
introduced for the Kerr metric \cite{bl}. Actually, the relevant metric
components are characterized by two integers $p$ and $q$ that describe the
rate with which these components approach zero (see below). Thus our main
concern is quite practical. Given $p$ and $q$, one may ask, for which
relations between them the metric near the horizon is regular and for which
is not. Also note, that this analysis does not depend on the type of gravity
theory.

As far as the notion of regularity is concerned, it is relevant in our
context in two aspects. (i) We require the finiteness of curvature
invariants near the horizon. (ii) We require that separate components of the
Riemann tensor, calculated in the frame, attached to a free falling
observer, be finite. Both requirements are not equivalent, (ii) is more
tight than (i). It is worth reminding that a stationary frame becomes
singular near the horizon even in the simplest Schwarzschild case since the
force and acceleration of an observer diverge. Meanwhile, a free falling
particle is free from this deficiency.

One reservation is in order. We do not discuss in the present paper naked 
\cite{nk1}, \cite{nk2} or truly naked \cite{truly} horizons, restricting to
usual regular ones.

The paper is organized as follows. In Sec. II we write the general form of
the metric under study and give explicit expression for the Riemann
curvature in terms of the 3+1 foliation. In Sec. III, IV and V we
investigate regularity of scalar invariants for nonextremal, extremal and
ultraextremal horizons, finding constraints on the near-horizon behavior of
metric coefficients. In Section VI we investigate regularity of the Riemann
tensor components in the tetrad attached to a free falling frame. All
results are generalized in Tables 1 and 2. In Appendix A we consider a
free-falling particle in the frame of zero-angular momentum observers (ZAMO) 
\cite{72} and find near-horizon behavior of different tetrad components of
the three-velocity. In Appendices B and C we give some mathematical details
needed for consideration of the near-horizon behavior of the geometry.

\section{General setup}

Let us consider the metric%
\begin{equation}
ds^{2}=-N^{2}dt^{2}+g_{\phi }(d\phi -\omega dt)^{2}+\frac{dr^{2}}{A}%
+g_{\theta }d\theta ^{2}\text{.}  \label{met}
\end{equation}

Here it is assumed that all coefficient may depend on $r$ and $\theta $
only. Hereafter, we use notations $g_{\theta }\equiv g_{\theta \theta }$ and 
$g_{\phi }\equiv g_{\phi \phi }$ for brevity.

Then, one can find that the Riemann curvature

\begin{eqnarray}
R &=&^{(3)}R+\dfrac{g_{\phi }}{2N^{2}}[A(\partial _{r}\omega )^{2}+g_{\theta
}^{-1}(\partial _{\theta }\omega )^{2}]-\Big[2\Big(A\dfrac{\partial _{r}^{2}N%
}{N}+g_{\theta }^{-1}\dfrac{\partial _{\theta }^{2}N}{N}\Big)+A\partial
_{r}(\gamma _{\phi }+\gamma _{\theta })\dfrac{\partial _{r}N}{N}
\label{R_exp} \\
&&+g_{\theta }^{-1}\partial _{\theta }(\gamma _{\phi }-\gamma _{\theta })%
\dfrac{\partial _{\theta }N}{N}+N^{-1}\Big(\partial _{r}A\partial
_{r}N-A^{-1}g_{\theta }^{-1}\partial _{\theta }A\partial _{\theta }N\Big)%
\Big].
\end{eqnarray}

Here, $^{(3)}R$ is the Riemann curvature of the submanifold $t=const$, $%
\gamma _{\theta }=\ln g_{\theta }$, $\gamma _{\phi }=\ln g_{\phi }$. Eq. (%
\ref{R_exp}) can be obtained, for example, from eq. (18) of \cite{tz}. It is
assumed that at $r=r_{g}$ there is a black hole horizon $N=0$.

In what follows, we consider separately the properties of horizons of
different kinds.

\section{Nonextremal horizon}

By definition, a notion of a nonextremal horizon means that the surface
gravity $\kappa $ is not equal to zero. Using the definition%
\begin{equation}
\kappa =\lim_{r\rightarrow r_{g}}\sqrt{\left( \nabla N\right) ^{2}},
\label{kappa}
\end{equation}%
we can write for our metric the expansion for the lapse function $N$ in the
form

\begin{equation}
N=a(\theta )\sqrt{2u}+\kappa _{1}(\theta )u+o(u),
\end{equation}%
where $u=r-r_{g}(\theta )$. The Ricci scalar in this case has the following
asymptotic form: 
\begin{equation}
R=\dfrac{1}{2}\Big(A(r,\theta )+\dfrac{r_{\theta }^{\prime }{}^{2}}{%
g_{\theta }}\Big)\dfrac{1}{u^{2}}+O\Big(\dfrac{1}{u}\Big).
\end{equation}

To make Ricci scalar to be divergent not faster than $1/u$ we have to choose 
$A(r,\theta )$ to be $O(u)$: $A(r,\theta )=A_{1}u+o(u)$ and $r_{g}(\theta
)=r_{g}=\mathrm{const}$, $A(r_{g},\theta )=0$.

Then, it turns out that 
\begin{equation}
R=\dfrac{g_{\phi }(\partial _{\theta }\omega )^{2}}{4g_{\theta \theta }a^{2}}%
\dfrac{1}{u}-\dfrac{\kappa _{2}}{2\sqrt{2}g_{\theta }a^{3}}(A_{1}g_{\theta
}a^{2}+g_{\phi }(\partial _{\theta }\omega )^{2})\dfrac{1}{\sqrt{u}}+O(1)%
\text{.}
\end{equation}

Elimination of these divergences may take place if we take $\kappa _{2}=0$
and 
\begin{equation}
\partial _{\theta }\omega \sim \sqrt{u}.  \label{do}
\end{equation}

Thus we have:\newline
\begin{equation}
N(r,\theta )=a(\theta )\sqrt{2u}+o(u),~~\omega (r,\theta )=\hat{\omega}%
_{H}+\omega _{1/2}\sqrt{u}+\omega _{2}u+o(u),  \label{N}
\end{equation}%
where $\hat{\omega}_{H}$ does not depend on $\theta .$ Hereafter $\hat{f}$
means that corresponding quantity $f$ is independent on $\theta $. Let us
consider now the traceless part of the Ricci tensor $Q_{\mu \nu }\equiv
R_{\mu \nu }-\frac{1}{4}Rg_{\mu \nu }$.

Then, $R_{2}\equiv 4Q_{\mu \nu }Q^{\mu \nu }=4R_{\mu \nu }R^{\mu \nu }-R^{2}$%
. We can find from the above expansions that%
\begin{equation}
R_{2}=\dfrac{11g_{\varphi \varphi }^{2}\partial _{\theta }\omega ^{4}}{%
16g_{\theta }^{2}a^{4}}\dfrac{1}{u^{2}}+O\Big(\dfrac{1}{u}\Big).
\end{equation}

Using (\ref{do}) we see that the leading divergences vanish. The $1/u$ term
may be written in a form: 
\begin{eqnarray}
\lim_{r\rightarrow r_{g}}(uR_{2}) &=&\dfrac{1}{4A_{1}g_{\theta \theta }a^{2}}%
[2a^{2}(\partial _{\theta }A_{1})^{2}+  \label{uR} \\
&&+8A_{1}a\partial _{\theta }A_{1}a_{\theta }^{\prime }+8A_{1}^{2}(a_{\theta
}^{\prime })^{2}]+\dfrac{11A_{1}g_{\varphi \varphi }^{2}}{8g_{\theta \theta
}a^{4}}(\partial _{\theta }\omega )^{2}(\partial _{r}\omega )^{2}.
\end{eqnarray}

For this limit to be equal to zero, we have to choose $\partial _{\theta
}A_{1}=\partial _{\theta }a=0$, giving thus expansions for $N$ and $A$ in
the form: 
\begin{equation}
N(r,\theta )=\hat{a}\sqrt{2u}+o(u),  \label{N1}
\end{equation}%
\begin{equation}
A(r,\theta )=\hat{A}_{1}u+o(u).  \label{A1}
\end{equation}%
Analyzing the last term, we see that we have to choose $\omega _{1/2}=0$,
thus an expansion for $\omega $ gives: 
\begin{equation}
\omega (r,\theta )=\hat{\omega}_{H}+\omega _{1}(\theta)u+o(u).  \label{om1}
\end{equation}

It follows from these expressions that 
\begin{eqnarray}
\lim_{r\rightarrow r_{g}}R &=&-\dfrac{A_{2}}{2}+\dfrac{\partial _{\theta
}\gamma _{\theta }\partial _{\theta }\gamma _{\varphi }+(\partial _{\theta
}\gamma _{\varphi })^{2}}{2g_{\theta }}-g_{\theta }^{-1}g_{\phi
}^{-1}\partial _{\theta }^{2}g_{\phi }-  \label{Rhor} \\
&&-\hat{A}_{1}\partial _{r}(\gamma _{\theta }+\gamma _{\varphi })-\dfrac{6%
\sqrt{2}\hat{A}_{1}\kappa _{3}}{\hat{a}}+\dfrac{1}{4}\dfrac{\hat{A}%
_{1}g_{\phi }\omega _{2}^{2}}{\hat{a}^{2}}.
\end{eqnarray}%
where $A_{2}$, $\kappa _{3}$ and $\omega _{2}$ are coefficients in the
expansions: 
\begin{equation}
N(r,\theta )=\hat{a}\sqrt{2u}+\kappa_{3/2}(\theta)u^{3/2}+O(u^{2})\text{,}
\label{Nau}
\end{equation}%
\begin{equation}
A(r,\theta )=\hat{A}_{1}u+A_{2}(\theta)u^{2}+o(u^{2}),  \label{A}
\end{equation}%
\begin{equation}
\omega (r,\theta )=\hat{\omega}_{H}+\omega _{1}(\theta)u+o(u).  \label{om}
\end{equation}

The fractional degrees of $u$ that appear in $N$, do not appear in the
metric coefficient $N^{2}$ in (\ref{met}). Actually, this is an expansion in
terms of $r-r_{g}$. We have to note that the surface gravity $\kappa =\hat{a}%
\sqrt{\dfrac{\hat{A}_{1}}{2}}$ also does not depend on $\theta $ as it
should be according to the zeroth law of black hole mechanics \cite{zero}.

The expression (\ref{Rhor}) is finite. Then, the finiteness of $\partial
_{r}\gamma _{\theta }$ and $\partial _{r}\gamma _{\phi }$ entails the \
finiteness of $\partial _{r}g_{\theta }$ and $\partial _{r}g_{\phi }$. Thus
the expansions for angular components of the metric read 
\begin{equation}
g_{\theta }=g_{\theta H}(\theta )+g_{\theta 1}(\theta )u+o(u)\text{,}
\label{g2}
\end{equation}%
\begin{equation}
g_{\phi }=g_{\varphi H}(\theta )+g_{\varphi 1}(\theta )u+o(u).  \label{g3}
\end{equation}

\section{Extremal horizon}

By definition of the extremal horizon, $\kappa =0$ and 
\begin{equation}
A=O(u^{2}).  \label{AE}
\end{equation}

Thus we have 
\begin{equation}
A(r,\theta )=A_{2}(r-r_{g}(\theta ))^{2}+o((r-r_{g}(\theta ))^{2}).
\end{equation}

This expansion gives the leading potentially divergent term $\,O(1/u^{2})$
in $R$ 
\begin{equation}
\lim_{r\rightarrow r_{g}(\theta )}(u^{2}R)=-\dfrac{4r_{g}^{\prime }(\theta
)^{2}}{g_{\theta \theta }}.
\end{equation}

To cancel this divergence, we have to choose $r_{g}=const$.

It turns out that this condition entails that the divergences $O(\frac{1}{u}%
) $ in $R$ and divergences $1/u^{2}$ and $1/u$ in $R_{2}$ also vanish. Now,
we will take into account in (\ref{R_exp}) the terms containing $A$ and $N$.
They read%
\begin{equation}
-\dfrac{\partial _{r}A\partial _{r}N+2A\partial _{r}^{2}N}{N}+g_{\theta
}^{-1}\dfrac{\partial _{\theta }A\partial _{\theta }N}{AN}.
\end{equation}

Their finiteness admits the asymptotic form 
\begin{equation}
N^{2}=O(u^{p}),  \label{NE}
\end{equation}%
with an arbitrary $p>0$. Along the direction with $\theta =const$ and $\phi
=const$, this is equivalent to $N^{2}\sim e^{-pn}$ in full analogy with the
spherically symmetric case ($n$ is the proper distance).

Now, we pay attention to the terms in (\ref{R_exp}) that equal 
\begin{equation}
\dfrac{g_{\phi }}{2N^{2}}(A(\partial _{r}\omega )^{2}+g_{\theta
}^{-1}(\partial _{\theta }\omega )^{2}).  \label{R_exp_main}
\end{equation}

To make these terms non-divergent and requiring that the expansion contain
only integer powers of $u$, we have to choose $\partial _{\theta }\omega
\sim N\sim O(u^{p/2})$ for even $p$, and $\partial _{\theta }\omega \sim
O(u^{(p+1)/2})$ for odd $p$, giving expansion for $\omega $.

In both cases (when $p=2k$ is even or $p=2k-1$ is odd), we can write
expansion

\begin{equation}
\omega =\hat{\omega}_{H}+\hat{\omega}_{1}u+..+\hat{\omega}_{k-1}u^{k-1}+\hat{%
\omega}_{k}(\theta )u^{k}+o(u^{k}),
\end{equation}%
where $\{\hat{\omega}_{H},\hat{\omega}_{1},...,\hat{\omega}_{k-1}\}$ are
independent on $\theta $. Equivalently, we may write 
\begin{equation}
\omega =\hat{\omega}_{H}+\hat{\omega}_{1}u+..+\hat{\omega}_{[\frac{p+1}{2}%
]-1}u^{[\frac{p+1}{2}]-1}+\hat{\omega}_{[\frac{p+1}{2}]}(\theta )u^{[\frac{%
p+1}{2}]}+o(u^{[\frac{p+1}{2}]}),
\end{equation}%
where $[...]$ means an integer part.

However, if we try to substitute this expansion in the first term in (\ref%
{R_exp_main}), the finiteness of $R$ requires that all the coefficients $%
\hat{\omega}_{s}$ with $s=1,2...k-1$ should to be equal to zero.

Thus

\begin{equation}
\omega =\hat{\omega} _{H}+\hat{\omega}_{k}(\theta )u^{k}+o(u^{k})\text{.}
\end{equation}

The expansions for $g_{\theta }$ and $g_{\phi }$ (\ref{g2}), (\ref{g3})
remain valid because they rely on the fact that $\gamma _{\theta }$ and $%
\gamma _{\phi }$ are finite in (\ref{R_exp}).

\section{Ultraextremal horizon}

Now we assume, similarly to the previous section, that $N^{2}$ has the order 
$u^{p}$. This means that 
\begin{equation}
N^{2}=\kappa _{p}(\theta )u^{p}+\kappa _{p+1}(\theta )u^{p+1}+o(u^{p+1})%
\text{.}  \label{Np}
\end{equation}

However, instead of $A\sim u^{2}$, typical of the extremal case in the
previous section, we consider the expansion for $A$ in a more general form 
\begin{equation}
A(r,\theta )=A_{q}(\theta)u^{q}+A_{q+1}(\theta)u^{q+1}+o(u^{q+1}),  \label{s}
\end{equation}%
with an integer $q>0$. By definition, the combination of (\ref{Np}) and (\ref%
{s}) gives us what is called "ultraextremal horizons".

Again, we require the finiteness of the Riemann curvature $R$, given in (\ref%
{R_exp}). First of all, we have to note that $^{(3)}R$ term does not involve 
$N.$ Let us start with the term 
\begin{equation}
\dfrac{g_{\phi }}{2N^{2}}[A(\partial _{r}\omega )^{2}+g_{\theta \theta
}^{-1}(\partial _{\theta }\omega )^{2}].  \label{term}
\end{equation}

It should be finite. Starting with a second term here, we see that we have
to choose $(\partial _{\theta }\omega )^{2}\sim O(u^{p})$, and taking into
account that expansion for $\omega $ has to be present only with integer
degrees of $u$, we come up with the expansion: 
\begin{equation}
\omega (r,\theta )=\hat{\omega} _{H}+\hat{\omega}_{1}u+...+\hat{\omega}_{[%
\frac{p-1}{2}]}u^{[\frac{p-1}{2}]}+\omega _{[\frac{p+1}{2}]}(\theta )u^{[%
\frac{p+1}{2}]}+\omega _{[\frac{p+3}{2}]}(\theta )u^{[\frac{p+3}{2}]}+...~,
\label{om_exp_frst}
\end{equation}%
where $\{\hat{\omega}_{1},...,\hat{\omega}_{[\frac{p-1}{2}]}\}$ do not
depend on $\theta $.

If $p=2k$,%
\begin{equation}
\omega (r,\theta )=\hat{\omega} _{H}+\hat{\omega}_{1}u+...+\hat{\omega}%
_{k-1}u^{k-1}+\omega _{k}(\theta )u^{k}+\omega _{k+1}(\theta )u^{k+1}+...~.
\label{p2q}
\end{equation}

If \thinspace $p=2k-1$, eq. (\ref{p2q}) is still valid.

Let the terms $\hat{\omega}_{i}=0$ for all $i=1,2,...s$, so the expansion
for $\omega $ starts from the term $O(u^{s+1})$. Taking also (\ref{s}) into
account, we have the condition $q\geq p-2s$, so 
\begin{equation}
A(r,\theta )=A_{p-2s}u^{p-2s}+A_{p-2s+1}u^{p-2s+1}+o(u^{p-2s+1}),
\label{apq}
\end{equation}%
where $p>2s$. For an even $p$, the maximum possible value $s=\frac{p}{2}-1$,
then $A\sim O(u^{2})$ and we return to the extremal case considered in the
previous Section. If $p=2l+1$ is odd, the maximum possible value $s=l$ and $%
A=O(u)$.

Analyzing other terms in (\ref{R_exp}) with $\partial _{r}N/N$ and $\partial
_{r}^{2}N/N$ 
\begin{equation}
A\Big(\partial _{r}\gamma _{\varphi }+\partial _{\theta }\gamma _{\theta }+%
\dfrac{\partial _{r}A}{A}-2\dfrac{\partial _{r}^{2}N}{N}\Big),
\end{equation}%
we see that their regularity leads either again to $A\sim O(u^{2})$, or to $%
N^{2}\sim A\sim u$. Other terms in (\ref{R_exp}) are proportional to $%
\partial _{\theta }N/N$ and $\partial _{\theta }^{2}N/N$, which lead to
restrictions for $\kappa _{n}$ or $r_{g}(\theta )$. Thus generalizing we have

\begin{tabular}{|p{0.25in}|p{1.5in}|p{2in}|p{2.25in}|}
\hline
& \textbf{Nonextremal} & \textbf{Extremal} & \textbf{Ultraextremal} \\ \hline
$N^{2}$ & $\hat{a}u+\kappa _{3}(\theta )u^{2}+o(u^{3})$ & $\kappa
_{p}u^{p}+o(u^{p})$ & $\kappa _{p}u^{p}+o(u^{p})$ \\ \hline
$A$ & $\hat{A}_{1}u+A_{2}(\theta )u^{2}+o(u^{2})$ & $A_{2}u^{2}+o(u^{2})$ & $%
A_{q}u^{q}+o(u^{q})$ \\ \hline
$\omega $ & $\hat{\omega}_{H}+\omega _{2}(\theta )u+o(u)$ & $\hat{\omega}%
_{H}+\hat{\omega}_{1}u+...+\hat{\omega}_{k-1}u^{k-1}+\omega _{k}(\theta
)u^{k}+o(u^{k})$ & $\hat{\omega}_{H}+\hat{\omega}_{l}u^{l}+...+\hat{\omega}%
_{k-1}u^{k-1}+\omega _{k}(\theta )u^{k}+o(u^{k})$ \\ \hline
$g_{ab}$ & $\left( g_{ab}\right) _{H}+\left( g_{ab}\right) _{1}u+o(u)$ & $%
\left( g_{ab}\right) _{H}+\left( g_{ab}\right) _{1}u+o(u)$ & $\left(
g_{ab}\right) _{H}+\left( g_{ab}\right) _{1}u+o(u)$ \\ \hline
\end{tabular}
\newline
TABLE 1: Table showing expansions of different metric coefficient for
nonextremal, extremal and ultraextremal horizons. Here $k=\Big[\dfrac{p+1}{2}
\Big],~l=\Big[\dfrac{p-q+3}{2}\Big]$.

\begin{itemize}
\item $s<[(p-1)/2]$. According to (\ref{apq}), $A\sim u^{p-2s}$. The proper
distance $dn\sim du/u^{p/2-s}$, whence $n\sim 1/u^{p/2-s-1}\rightarrow
\infty $, $u\sim n^{-\frac{1}{p/2-s-1}}$, thus $N^{2}\sim u^{p}\sim n^{-%
\frac{p}{p/2-s-1}}$.

\item $s=[(p-1)/2]$ case. If p is even, than this again leads to $A\sim
u^{2} $. In this case we have for a proper distance: $dn=\dfrac{du}{\sqrt{A}}%
\sim \dfrac{du}{u}$, hence $n\sim C\ln u$ ($C$ is some constant) and thus $%
N^{2}\sim u^{p}\sim e^{-\frac{p}{C}n}$. We have an extremal horizon which is
analyzed in the previous section. In case when $p$ is odd we have $A\sim u$
(which inevitably leads, as we showed above, to $N^{2}\sim u$). Thus we see
that we get no new restrictions as compared to those obtained from the
analysis behavior of $\omega $.
\end{itemize}

As far as the expansions of the quantities (\ref{g2}), (\ref{g3}) are
concerned, they retain their validity.

It is convenient to summarize the results listed above in Table 1. From this
table follows that required regularity of scalar invariants entail simple
conditions on $\omega $:

\begin{equation}  \label{om_R_reg}
\partial_{\theta}\omega=O(N),~~~\partial_r\omega=O(N/\sqrt{A}).
\end{equation}

\section{Tetrad components of the curvature tensor}

In the above consideration, we required the finiteness of some curvature
invariants. Meanwhile, physically, it is natural to demand something more.
Namely, not only the combination of the curvature components entering these
invariants should be finite, but also each component separately, if it is
measured in a proper frame. This means that a frame itself should not become
singular in contrast to the frame of a stationary observer since even in the
simplest case of the Schwarzschild metric such a frame becomes singular on
the horizon, the scalar of acceleration diverges.

The most natural choice is a tetrad attached to a free falling observer who
crosses the horizon without experiencing infinite tidal forces since the
geometry is regular there by definition.

\subsection{Orbital ZAMO frame (OZAMO)}

Now, we need to define a tetrad carried by an observer. For simplicity, it
is convenient to choose a zero angular momentum observer (ZAMO) \cite{72}.
Originally, they were introduced for observers orbiting around a black hole
(we call them OZAMOs). However, they are not free moving and, moreover,
become singular in the horizon limit. In the particular case of a static
black hole they represent static observers. Below, we will use the frame
composed of free-falling ZAMOs (FZAMO).

To begin with, we consider the OZAMO frame first and pass to the FZAMO frame
later.

Such a frame can be realized by the tetrad%
\begin{eqnarray}
h_{(0)\mu } &=&-N(1,0,0,0),~~~h_{(1)\mu }=\sqrt{g_{\phi }}(-\omega
,1,0,0),~~~  \label{ozamo} \\
h_{(2)\mu } &=&\dfrac{1}{\sqrt{A}}(0,0,1,0),~~~h_{(3)\mu }=\sqrt{g_{\theta }}%
(0,0,0,1).  \label{ozamo_fr_last}
\end{eqnarray}

In this frame the tetrad components%
\begin{equation}
\tilde{R}_{\alpha \beta \gamma \delta }\equiv R_{\mu \nu \rho \sigma
}h_{(\alpha )}^{\mu }h_{(\beta )}^{\nu }h_{(\gamma )}^{\rho }h_{(\delta
)}^{\sigma }\text{,}
\end{equation}%
where Greek indices run from $0$ to $3.$

Explicit expressions for the Riemann curvature tensor in this tetrad frame
can be found in \cite{chandr} (Sec. 6. 51, eq. (3)), where another notations
were used ($\nu =\ln N$, $2\psi =\ln g_{\phi }$, $2\mu _{2}=-\ln A$, $2\mu
_{3}=\ln g_{\theta }$). One can check directly that our conditions of
regularity of scalar curvature (\ref{om_R_reg}) make all the curvature
tensor components in this frame regular. This is consistent with a more
detailed information containing in Table I and also agrees with a similar
observation made in \cite{tz} (Section IV A) where somewhat different
coordinate system was used.

\subsection{Falling ZAMO frame (FZAMO)}

To get expansions for the curvature tensor in FZAMO frame, we choose a
tetrad that is attached to a free falling observer, where the temporal basis
vector is directed along its four-velocity and the other three are
orthogonal to it. This can be done in the following way.

\begin{itemize}
\item We choose tetrad of OZAMO frame (\ref{ozamo}),

\item rotate the frame in the $\theta $-$r$ plane by angle $\psi $: 
\begin{eqnarray}
\tilde{e}_{(2)} &=&h_{(2)}\cos \psi +h_{(3)}\sin \psi , \\
\tilde{e}_{(3)} &=&-h_{(2)}\sin \psi +h_{(3)}\cos \psi , \\
\tilde{e}_{(1)} &=&h_{(1)}~~~\tilde{e}_{(0)}=h_{(0)},
\end{eqnarray}

\item rotate the frame in the $\varphi $-$r$ plane by angle $\delta $: 
\begin{eqnarray}
\tilde{e}_{(2)}^{\prime}&=&\tilde{e}_{(2)}\cos \delta +\tilde{e}_{(1)}\sin
\delta , \\
\tilde{e}_{(1)}^{\prime}&=&-\tilde{e}_{(2)}\sin \delta +\tilde{e}_{(1)}\cos
\delta , \\
\tilde{e}_{(3)}^{\prime}&=&\tilde{e}_{(3)}~~~\tilde{e}_{(0)}=\tilde{e}_{(0)},
\end{eqnarray}

\item boost it in the radial direction: 
\begin{eqnarray}
\hat{e}_{(0)} &=&\gamma (\tilde{e}_{(0)}^{\prime }+\upsilon \tilde{e}%
_{(2)}^{\prime }),~~~~~\hat{e}_{(1)}=\tilde{e}_{(1)}^{\prime }, \\
\hat{e}_{(2)} &=&\gamma (\tilde{e}_{(2)}^{\prime }+\upsilon \tilde{e}%
_{(0)}^{\prime }),~~~~~\hat{e}_{(3)}=\tilde{e}_{(3)}^{\prime },
\end{eqnarray}%
where $\gamma =E/N$ and $\upsilon =\sqrt{1-1/\gamma ^{2}}.$
\end{itemize}

We also impose an additional conditions: $\psi =O(N)$ and $\delta=O(N)$ (see
Appendix A). We will use notations%
\begin{equation}
\hat{R}_{\alpha \beta \gamma \delta }\equiv R_{\mu \nu \rho \sigma }\hat{e}%
_{(\alpha )}^{\mu }\hat{e}_{(\beta )}^{\nu }\hat{e}_{(\gamma )}^{\rho }\hat{e%
}_{(\delta )}^{\sigma }\text{.}
\end{equation}

Corresponding relations which occur from regularity of $\hat{R}_{\alpha
\beta \gamma \delta }$ are listed in Appendix B

\subsubsection{Nonextremal horizon}

Using the expansions for nonextremal horizons (\ref{Nau} - \ref{g3}) we
obtain near the horizon

\begin{equation}
\hat{R}_{0101}=-\dfrac{3E^{2}\hat{A}_{1}}{16a^{2}g_{\phi H}}\dfrac{\hat{g}%
_{\phi 3/2}}{\sqrt{u}}+O(1),
\end{equation}%
\begin{equation}
\hat{R}_{0202}=O(1),
\end{equation}%
\begin{equation}
\hat{R}_{0303}=-\dfrac{3E^{2}\hat{A}_{1}}{16a^{2}g_{\theta H}}\dfrac{\hat{g}%
_{\theta 3/2}}{\sqrt{u}}+O(1),
\end{equation}%
\begin{equation}
\hat{R}_{0102}=-\dfrac{3E}{16a^{2}}\sqrt{g_{\phi H}}\dfrac{\hat{\omega}_{3/2}%
}{\sqrt{u}}+O(1),
\end{equation}%
\begin{equation}
\hat{R}_{0103}=\dfrac{E^{2}}{16\sqrt{2}a^{3}}\sqrt{\dfrac{g_{\phi H}}{%
g_{\theta H}}}\Big(3\hat{A}_{1}\sqrt{g_{\theta H}}\dfrac{\psi }{\sqrt{u}}%
\hat{\omega}_{3/2}-2\sqrt{A}_{1}\hat{\omega}_{3/2}^{\prime }\Big)+O(1),
\end{equation}%
where $\hat{a}$ and $\hat{A}_{1}$ are coefficients in expansions (\ref{Nau})
and (\ref{A}). The conditions of boundness of these expressions are: 
\begin{equation}
\hat{g}_{\theta 3/2}=\hat{g}_{\phi 3/2}=\hat{\omega}_{3/2}=0.
\end{equation}%
This means that the main terms in the expansions of these quantities contain
only integer degrees of $u$: 
\begin{gather}
g_{a}=g_{Ha}(\theta )+g_{1a}(\theta )u+O(u^{2}),~~a=\varphi ,\theta , \\
\omega =\hat{\omega}_{H}+\omega _{1}(\theta )u+O(u^{2}).
\end{gather}

\subsubsection{Extremal horizon}

In this subsection we consider the properties of the metric when $q=2$. In
this case we have expansions in the form: 
\begin{gather}
A=A_{2}(\theta )u^{2}+A_{3}(\theta )u^{3}+o(u^{3}), \\
N^{2}=\kappa _{p}(\theta )u^{p}+\kappa _{p+1}(\theta )u^{p+1}+o(u^{p+1}), \\
\omega =\hat{\omega} _{H}+\hat{\omega}_{k}u^{k}+...+\hat{\omega}%
_{n-1}u^{n-1}+\omega _{n}(\theta )u^{n}+o(u^{n}), \\
k=\Big[\dfrac{p-q+3}{2}\Big],~~~n=\Big[\dfrac{p+1}{2}\Big], \\
g_{a}=g_{aH}(\theta )+g_{a1}(\theta )u+o(u),~~~a=\varphi ,\theta .
\end{gather}

We will start from the $\hat{R}_{0313}$ component, given by (\ref{R_0313_rel}%
) , where%
\begin{equation}
\partial _{\theta }(\gamma _{\theta }-3\gamma _{\varphi }+\ln N^{2})\partial
_{\theta }\omega -2\partial _{\theta }^{2}\omega =O(N^{2}).  \label{om_eq}
\end{equation}

Substituting given expansions, we have an equation for the $u^{0}$ term on
the left hand side of this equation: 
\begin{equation}
\partial _{\theta }\omega _{n}\partial _{\theta }\ln \dfrac{\kappa
_{p}g_{\theta H}}{g_{\varphi H}^{3}}=2\partial _{\theta }^{2}\omega
\rightarrow ~\partial _{\theta }\omega _{n}=C\sqrt{\dfrac{\kappa
_{p}g_{\theta H}}{g_{\varphi H}^{3}}},~~C=\mathrm{const.}
\end{equation}

We will assume that the space-time has no conical defects, so for $\theta
\rightarrow 0$ or $\theta \rightarrow \pi \,\ $the coefficient $g_{\varphi
H}\sim \sin ^{2}\theta $. Then, to have regular behavior of $\omega _{n}$,
we take $C=0$.

Taking now the $u^{1}$ term from (\ref{om_eq}), we have the same equation
for $\omega _{n+1}^{\prime }$ that gives $\omega _{n+1}^{\prime }=0$ as a
solution. This will continue up to the $\omega _{p}$ term, and we get an
expansion for $\omega $: 
\begin{equation}  \label{omega_exp}
\omega =\hat{\omega} _{H}+\hat{\omega}_{k}u^{k}+...+\hat{\omega}%
_{p-1}u^{p-1}+\omega _{p}(\theta )u^{p}+o(u^{p}),
\end{equation}%
where hat over some coefficients $\hat{\omega}_{i}$ means that $\hat{\omega}%
_{i}=const.$

The conditions for $A$ and $N^{2}$ can be obtained from $\hat{R}_{0101}$ and 
$\hat{R}_{0303}$. These relations are considered in subsection \ref{ue},
where the $p\neq q$ case is considered. The solution is given by (\ref{A_rel}%
), where we have to choose $q=2.$

The analysis of the components $\hat{R}_{0203}$ and $\hat{R}_{0113}$ give no
additional constraints on the metric coefficients. If we assume that $%
\psi\sim N$ (see. Appendix A), $\hat{R}_{0103}$ gives an additional
conditions for $\omega $: $\partial _{\theta }\omega \sim u^{p-1}u^{\frac{p+2%
}{2}}$. So, our expansions have the form: 
\begin{gather}
\omega =\hat{\omega} _{H}+\hat{\omega}_{k}u^{k}+...+\omega _{3p/2}(\theta
)u^{3p/2}+o(u^{3p/2}), \\
A=A_{2}u^{2}+o(u^{2})~~~N^{2}=\kappa _{p}u^{p}+o(u^{p}), \\
g_{a}=g_{aH}(\theta )+g_{a,p}(\theta )u^{p}+o(u^{p}),~~~a=\varphi ,\theta .
\end{gather}

In the particular case $p=2$ we have for $k=1$, $n=1,$%
\begin{equation}
\omega =\hat{\omega} _{H}+\hat{\omega}_{1}u+\omega _{2}(\theta
)u^{2}+o(u^{2}).
\end{equation}

It is instructive to give explicit formulas for the case $p=2$. Then, we can
find expressions for the curvature tensor in the FZAMO frame: 
\begin{eqnarray}
\hat{R}_{0101} &\sim &\hat{R}_{1212}\sim \hat{R}_{0112}\sim \dfrac{1}{u^{2}}%
\Big[\Big(\dfrac{A_{2}^{\prime }}{A_{2}}+\dfrac{\kappa _{2}^{\prime }}{%
\kappa _{2}}\Big)g_{\varphi H}^{\prime }+g_{\varphi H}^{2}\dfrac{\omega
_{1}^{\prime }{}^{2}}{\kappa _{2}}\Big], \\
\hat{R}_{0103} &\sim &\hat{R}_{0123}\sim \hat{R}_{0312}\sim \hat{R}%
_{1223}\sim \dfrac{1}{u^{2}}\Big[2\omega _{1}^{\prime }-\Big(\dfrac{%
A_{2}^{\prime }}{A_{2}}+\dfrac{\kappa _{2}^{\prime }}{\kappa _{2}}\Big)%
\omega _{1}\Big], \\
\hat{R}_{0203} &\sim &\hat{R}_{0223}\sim \dfrac{1}{u}\Big[\dfrac{\kappa
_{2}^{\prime }}{\kappa _{2}}+\dfrac{A_{2}^{\prime }}{A_{2}}\Big], \\
\hat{R}_{0113} &\sim &\dfrac{\omega _{1}\omega _{1}^{\prime }}{u}.
\end{eqnarray}

All these expansions lead to the conditions: $\omega _{1}^{\prime }=\kappa
_{2}^{\prime }=A_{2}^{\prime }=0$.

Under these conditions the expressions for $\hat{R}_{\mu \nu \rho \sigma }$
are further simplified and we see, in particular, that 
\begin{equation}
\hat{R}_{0101}\sim \hat{R}_{0103}\sim \hat{R}_{1212}\sim \hat{R}_{0112}\sim 
\dfrac{1}{u}\Big(\dfrac{\kappa _{3}^{\prime }}{\kappa _{2}}+\dfrac{%
A_{3}^{\prime }}{A_{2}}\Big).
\end{equation}

Thus we have $\omega _{1}^{\prime }=\kappa _{2}^{\prime }=A_{2}^{\prime
}=\kappa _{3}^{\prime }=A_{3}^{\prime }=0$. They can be rewritten as 
\begin{equation}
\partial _{\theta }\omega =O(N^{2}),~~~~~~~\dfrac{\partial _{\theta }N^{2}}{%
N^{2}}=O(N^{2}),~~~~~~~\dfrac{\partial _{\theta }A^{2}}{A^{2}}=O(A^{2}).
\end{equation}

The results of this subsection agree with Sec. IVB3 of \cite{tz}.

\subsubsection{Ultraextremal horizon\label{ue}}

In this subsection we consider the properties of metric in the case of the
ultraextremal horizon.

\paragraph{\textbf{Simplest case: $p=q$}.}

We will start from the simplest case, when $p=q$. Then, general relations,
given by (\ref{comp_frs}-\ref{lst_conf}), become: 
\begin{gather}  
\hat{R}_{0113}:\dfrac{\partial _{\theta }\omega \partial _{r}\omega }{N^{2}}%
=O(1),  \label{r0101_p=q} \\
\hat{R}_{0101}:\Big(\dfrac{\partial _{\theta }A}{A}+\dfrac{\partial _{\theta
}N^{2}}{N^{2}}\Big)\partial _{\theta }\gamma _{\varphi }+g_{\varphi \varphi }%
\Big(\dfrac{\partial _{\theta }\omega }{N}\Big)^{2}=O(N^{2}), \\
\hat{R}_{0203}:\dfrac{\partial _{\theta }A}{A}\dfrac{\partial _{r}N^{2}}{%
N^{2}}+2\dfrac{\partial _{r}\partial _{\theta }N^{2}}{N^{2}}-\dfrac{\partial
_{\theta }N^{2}}{N^{2}}\dfrac{\partial _{r}N^{2}}{N^{2}}=O(1), \\
\hat{R}_{0303}:\Big(\dfrac{\partial _{\theta }A}{A}+\dfrac{\partial _{\theta
}N^{2}}{N^{2}}\Big)\partial _{\theta }\gamma _{\theta }+3\Big(\dfrac{%
\partial _{\theta }A}{A}\Big)^{2}+\Big(\dfrac{\partial _{\theta }N^{2}}{N^{2}%
}\Big)^{2}+ \\
+3g_{\varphi \varphi }\Big(\dfrac{\partial _{\theta }\omega }{N}\Big)^{2}-2%
\Big(\dfrac{\partial _{\theta }^{2}A}{A}+\dfrac{\partial _{\theta }^{2}N^{2}%
}{N^{2}}\Big)=O(N^{2}),\label{r_0303_q=p} \\
\hat{R}_{0313}:\partial _{\theta }(\gamma _{\theta }-3\gamma _{\varphi
})\partial _{\theta }\omega +\dfrac{\partial _{\theta }N^{2}}{N^{2}}\partial
_{\theta }\omega -2\partial _{\theta }^{2}\omega =O(N^{2}), \\
\hat{R}_{0103}:\partial _{\theta }\omega \partial _{r}(\gamma _{\theta
}-3\gamma _{\varphi })+2\partial _{\theta }\omega \dfrac{\partial _{r}N^{2}}{%
N^{2}}-\Big(\dfrac{\partial _{\theta }A}{A}+\dfrac{\partial _{\theta }N^{2}}{%
N^{2}}\Big)\partial _{r}\omega -2\partial _{r}\partial _{\theta }\omega
=O(N^{2}).
\end{gather}

The relation that follows from the expression for $\hat{R}_{0313}$ is the
same as before, and the whole analysis, given after (\ref{om_eq}) is
relevant. This gives us $\partial _{\theta }\omega =O(N^{2})$. Taking this
into account, we see that $\hat{R}_{0101}$ gives the relation (if $g_{a}$
are given by general expansions (\ref{g2}-\ref{g3}))%
\begin{equation}
\Big(\dfrac{\partial _{\theta }A}{A}+\dfrac{\partial _{\theta }N^{2}}{N^{2}}%
\Big)=O(N^{2}).  \label{A_N2_cond_1}
\end{equation}

This relation is studied in Appendix C. The final result is given by (\ref%
{A_rel}), but there we have to choose $p=q$. The relation from $\hat{R}%
_{0203}$ gives us $A_{p}^{\prime }/A_{p}+\kappa _{p}^{\prime }/\kappa _{p}=0$%
which is consistent with (\ref{A_rel}). Now let us turn to $\hat{R}_{0303}$.
The corresponding relation takes the form: 
\begin{equation}
3\Big(\dfrac{\partial _{\theta }A}{A}\Big)^{2}+\Big(\dfrac{\partial _{\theta
}N^{2}}{N^{2}}\Big)^{2}-2\Big(\dfrac{\partial _{\theta }^{2}A}{A}+\dfrac{%
\partial _{\theta }^{2}N^{2}}{N^{2}}\Big)=O(N^{2}).
\end{equation}

Now we want to rewrite $\Big(\dfrac{\partial _{\theta }^{2}A}{A}+\dfrac{%
\partial _{\theta }^{2}N^{2}}{N^{2}}\Big)$ in terms of the first derivatives
by $\theta $. If we differentiate equation (\ref{A_N2_cond_1}) by $\theta $,
we obtain 
\begin{equation}
\dfrac{\partial _{\theta }^{2}A}{A}+\dfrac{\partial _{\theta }^{2}N^{2}}{%
N^{2}}-\Big(\dfrac{\partial _{\theta }A}{A}\Big)^{2}-\Big(\dfrac{\partial
_{\theta }N^{2}}{N^{2}}\Big)^{2}=O(N^{2}).
\end{equation}

We can rewrite this in the form 
\begin{equation}
\dfrac{\partial _{\theta }^{2}A}{A}+\dfrac{\partial _{\theta }^{2}N^{2}}{%
N^{2}}=\Big(\dfrac{\partial _{\theta }A}{A}\Big)^{2}+\Big(\dfrac{\partial
_{\theta }N^{2}}{N^{2}}\Big)^{2}+O(N^{2}).
\end{equation}

Substituting this in (\ref{A_N2_cond_1}), we have 
\begin{equation}
3\Big(\dfrac{\partial _{\theta }A}{A}\Big)^{2}+\Big(\dfrac{\partial _{\theta
}N^{2}}{N^{2}}\Big)^{2}-2\Big(\dfrac{\partial _{\theta }A}{A}\Big)^{2}-2\Big(%
\dfrac{\partial _{\theta }N^{2}}{N^{2}}\Big)^{2}=O(N^{2})\rightarrow
\end{equation}%
\begin{equation}
\rightarrow \Big(\dfrac{\partial _{\theta }A}{A}\Big)^{2}-\Big(\dfrac{%
\partial _{\theta }N^{2}}{N^{2}}\Big)^{2}=O(N^{2})\rightarrow \Big(\dfrac{%
\partial _{\theta }A}{A}+\dfrac{\partial _{\theta }N^{2}}{N^{2}}\Big)\Big(%
\dfrac{\partial _{\theta }A}{A}-\dfrac{\partial _{\theta }N^{2}}{N^{2}}\Big)%
=O(N^{2}).
\end{equation}

Taking into account (\ref{A_N2_cond_1}) and a fact, that expression inside
the second bracket has the order $O(1)$, we see that the condition of
regularity for $\hat{R}_{0303}$ is satisfied. It is easy to check that all
other components are regular as well.

Now, we want to formulate the corresponding conditions of regularity in
terms of the metric expansion. It is quite hard to analyze a general case,
so we will present only the case when $p=q=3$. Here, the following variants
are possible.

\begin{itemize}
\item First variant gives us%
\begin{equation}
\kappa _{3}^{\prime }=\kappa _{4}^{\prime }=\kappa _{5}^{\prime
}=0,~~~~~\omega _{H}^{\prime }=\omega _{1}^{\prime }=\omega _{2}^{\prime }=0.
\end{equation}%
Expansions for $g_{\phi }$ and $g_{\theta }$ are given by general
expressions (\ref{g2}-\ref{g3}).

\item Second variant: 
\begin{eqnarray}
\kappa _{3}^{\prime } &=&\kappa _{4}^{\prime }=0~but~\kappa _{3}\neq
0,~~~~~\omega _{H}^{\prime }=\omega _{2}^{\prime }=0,~~~~~\omega _{1}=0, \\
g_{\varphi H}^{\prime } &=&0~~~~~g_{\theta H}=C_{1}(\kappa _{5}^{\prime
})^{2}.
\end{eqnarray}

\item Third variant: 
\begin{eqnarray}
\kappa _{3}^{\prime } &=&0~but~\kappa _{4}^{\prime }\neq 0,~~~~~\omega
_{H}^{\prime }=0,~~~~\omega _{1}=\omega _{2}=0, \\
g_{\phi H}^{\prime } &=&g_{\phi 1}^{\prime }=0, \\
g_{\theta H} &=&C_{1}\Big(\dfrac{\kappa _{4}^{\prime }}{\kappa _{3}}\Big)%
^{2},~~~g_{\theta 1}=2g_{\theta H}\Big(\dfrac{\kappa _{5}^{\prime }}{\kappa
_{4}^{\prime }}-\dfrac{\kappa _{4}}{\kappa _{3}}+C_{2}\Big).
\end{eqnarray}
\end{itemize}

The condition $\kappa _{3}^{\prime }=0$ comes from the regularity of $\hat{R}%
_{0203}$. The conditions for $\omega _{1}$ or $\omega _{2}^{\prime }$ come
from $\hat{R}_{0113}$. The conditions for $\kappa _{4}^{\prime }$ and $%
g_{\varphi H}^{\prime }$ or $g_{\varphi 1}^{\prime }$ and $g_{\varphi
H}^{\prime }$ come from $\hat{R}_{0101}$. The conditions for $g_{\theta H}$
and $g_{\theta 1}$ come from $\hat{R}_{0303}$.

These results agree with those from Sec. IV B 5 of \cite{tz}.

\paragraph{\textbf{General case: $p\neq q$}.}

Now let us consider the case when $p\neq q$. Then, for $q>p$ condition (\ref%
{om_eq}) obtained from regularity of $\hat{R}_{0313}$ (\ref{R_0313_rel}) is
valid. In this case we have the same condition $\partial _{\theta }\omega
\sim N^{2}$. If $q<p,$ it is very hard to derive the general regularity
condition explicitly. Therefore, we consider only a particular case. We
assume that in the condition from $\hat{R}_{0313}$ (\ref{R_0313_rel}) the
first 3 terms and the last one have the order $N^{2}$ independently. In this
case we obtain (\ref{omega_exp}) and, as a result, $\partial _{\theta
}\omega \sim N^{2}$, but with an additional constraint 
\begin{equation}
\partial _{r}\gamma _{\theta }\sim N/\sqrt{A}.  \label{gamma_cond_1}
\end{equation}

To get a relation for $A$ and $N^{2}$, we will analyze $\hat{R}_{0101}$ and $%
\hat{R}_{0303}$. This gives us: 
\begin{gather}
\Big(\dfrac{\partial _{\theta }A}{A}+\dfrac{\partial _{\theta }N^{2}}{N^{2}}%
\Big)\partial _{\theta }\gamma _{\varphi }-g_{\theta \theta }A\Bigg(\Big(%
\dfrac{\partial _{r}A}{A}-\dfrac{\partial _{r}N^{2}}{N^{2}}\Big)\partial
_{r}\gamma _{\varphi }+(\partial _{r}\gamma _{\varphi })^{2}+2\partial
_{r}^{2}\gamma _{\varphi }\Bigg)+\sqrt{A}\cdot O(\psi )=O(N^{2}),
\label{A_N2_rel_gen} \\
\Big(\dfrac{\partial _{\theta }A}{A}+\dfrac{\partial _{\theta }N^{2}}{N^{2}}%
\Big)\partial _{\theta }\Big(\gamma _{\theta }+\ln \dfrac{A}{N^{2}}\Big)%
-g_{\theta \theta }A\Bigg(\Big(\dfrac{\partial _{r}A}{A}-\dfrac{\partial
_{r}N^{2}}{N^{2}}\Big)\partial _{r}\gamma _{\theta }+(\partial _{r}\gamma
_{\theta })^{2}+2\partial _{r}^{2}\gamma _{\theta }\Bigg)+ \\
+\sqrt{A}\cdot O(\psi )=O(N^{2}).  \label{A_N2_rel_gen_2}
\end{gather}%

In these relations we denoted as $O(\psi )$ all terms, proportional to $\psi
.$ Note that if $q>p$, both these relations give (\ref{A_N2_cond_1}) (in
doing so, the terms proportional to $\psi $ are regular). If $q<p$, we can
find the regularity conditions for the metric coefficients in particular
cases only. When (\ref{A_N2_cond_1}) holds, the second term and terms of the
order $\psi $ in both equations (\ref{A_N2_rel_gen}) - (\ref{A_N2_rel_gen_2}%
) have order $N^{2}$. In this case, we can use the same solution, given by (%
\ref{A_rel}). Regularity of the second term gives us $\partial _{r}\gamma
_{a}\sim u^{p-q+1}$ which is stronger, than (\ref{gamma_cond_1}). As a
result, we obtain an expansion for the angular metric coefficients: 
\begin{equation}
g_{a}=g_{aH}+g_{a,p-q+2}u^{p-q+2}+o(u^{p-q+2}),~~a=\varphi ,\theta .
\label{g_exp_spec}
\end{equation}

Because of this expansion, terms containing $\psi $ turn out to be regular.

If the expansions (\ref{g_exp_spec}), (\ref{A_rel}) and (\ref{om_eq}) of
metric coefficients are valid, all components of the curvature tensor are
regular, possibly except from $\hat{R}_{0103}$ and $\hat{R}_{0203}$.
Regularity of $\hat{R}_{0103}$ leads to additional requirements $\partial
_{\theta }\omega \sim u^{3p/2-q/2+1}$ and $\partial _{r}\omega \sim
u^{p-q+1} $. These conditions and (\ref{g_exp_spec}) may be used to rewrite
the one for $\hat{R}_{0203}$: 
\begin{equation}
\dfrac{\sqrt{A}}{N}\Big[\Big(\dfrac{\partial _{\theta }A}{A}-\dfrac{\partial
_{\theta }N^{2}}{N^{2}}\Big)\dfrac{\partial _{r}N^{2}}{N^{2}}+2\dfrac{%
\partial _{r}\partial _{\theta }N^{2}}{N^{2}}\Big]=O(1).
\end{equation}

Using (\ref{A_rel}), this gives us 
\begin{equation}
2\dfrac{\partial _{r}\partial _{\theta }N^{2}}{N^{2}}-2\dfrac{\partial
_{r}N^{2}}{N^{2}}\dfrac{\partial _{\theta }N^{2}}{N^{2}}=O\Big(\dfrac{N}{%
\sqrt{A}}\Big).
\end{equation}

It follows from it that $\partial _{r}\Big(\dfrac{\partial _{\theta }N^{2}}{%
N^{2}}\Big)\sim \dfrac{N}{\sqrt{A}}$. Integrating over $r$, we obtain $%
\dfrac{\partial _{\theta }N^{2}}{N^{2}}\sim u^{\frac{p-q}{2}+1}$. In terms
of the expansion coefficients this can be rewritten as 
\begin{equation}
\kappa _{p}^{\prime }=...=\kappa _{p+\frac{p-q}{2}}^{\prime }=0.
\end{equation}

Using (B18), this leads to the corresponding condition for $A$: 
\begin{equation}
A_{q}^{\prime }=...=A_{q+\frac{p-q}{2}}^{\prime }=0.
\end{equation}

Thus summarizing, we have 
\begin{equation}
\begin{cases}
~\kappa _{p+l}^{\prime }=0,~A_{q+l}^{\prime }=0~\mathrm{for}~0\leq l\leq 
\dfrac{p-q}{2}, \\ 
~A~\mathrm{and}~N^{2}~\mathrm{are~related~by~(\ref{A_rel})~for}~\dfrac{p-q}{2%
}<l<p, \\ 
~\mathrm{No~special~condition~for}~l\geq p.%
\end{cases}\label{A_N^2_rel}
\end{equation}

{
\begin{tabular}{|p{1.7in}|p{1.25in}|p{1.5in}|p{1.5in}|}
\hline
\textbf{What is bounded} & \textbf{Nonextr. $q=1$} & \textbf{Extr. $q=2$} & 
\textbf{Ultraextr. $q>2$} \\ \hline
\textbf{Ricci scalar} &  &  &  \\ \hline
$p=q$ & $\omega _{H}^{\prime }=0$ & $\omega _{H}^{\prime }=0$ & $\omega
_{H}^{\prime }=...=\omega _{k-1}^{\prime }=0$ \\ \hline
$p>q$ & Singular & $\omega _{H}^{\prime }=0,~\omega _{1}=\omega _{k-1}=0$ & $%
\omega _{H}^{\prime }=0,~\omega _{1}=\omega _{l-1}=0,~\omega _{l}^{\prime
}=..=\omega _{k-1}^{\prime }=0$ \\ \hline
$p<q$ & Singular & Singular & $\omega _{H}^{\prime }=...=\omega
_{k-1}^{\prime }=0$ \\ \hline
\textbf{Quadratic invariants} &  &  &  \\ \hline
$p=q$ & $\kappa _{1}^{\prime }=A_{1}^{\prime }=0$ & ... & ... \\ \hline
$p>q$ & Singular & ... & ... \\ \hline
$p<q$ & Singular & Singular & ... \\ \hline
\textbf{Curv. tensor in FZAMO frame} &  &  &  \\ \hline
$p=q$ & ... & $\omega _{k}^{\prime }=...=\omega _{p-1}^{\prime }=0$ and (\ref%
{A_rel})* & $\omega _{k}^{\prime }=...=\omega _{p-1}^{\prime }=0$ and (\ref%
{A_rel})* \\ \hline
$p>q$ & Singular & $\omega _{k}^{\prime }=...=\omega _{p-1}^{\prime }=0$, $%
\partial _{r}^{2}g_{a}\sim N^{2}/A,~~~a=\theta ,~\varphi $ and (\ref%
{A_N^2_rel})* & $\omega _{k}^{\prime }=...=\omega _{p-1}^{\prime }=0$, $%
\partial _{r}^{2}g_{a}\sim N^{2}/A,~~~a=\theta ,~\varphi $ and (\ref%
{A_N^2_rel})* \\ \hline
$p<q$ & Singular & Singular & $\omega _{k}^{\prime }=...=\omega
_{p-1}^{\prime }=0$, $\partial _{r}g_{a}=O(1)$ and (\ref{A_rel})* \\ \hline
\end{tabular}%
}\\
{~~~~~$...$ means that condition is the same as in previous row for
the same relation between $p$ and $q$. (\ref{A_rel})* means that
coefficients in expansions are related by equation (\ref{A_rel}) from
appendix C, (\ref{A_N^2_rel})* means that relation is given by eq. (\ref%
{A_N^2_rel}). Each condition in each row means that a corresponding
condition coincides with the corresponding one in the previous row with the
same values of $p$ and $q$. The case $q<p$ is given  with reservations made
in the main text before eq. (\ref{gamma_cond_1}). } 
TABLE 2: Table, showing what conditions on metric coefficients are imposed
by Ricci scalar, Quadratic invariants and Curvature tensor in FZAMO frame.
Here $k=\Big[\dfrac{p+1}{2}\Big],~l=\Big[\dfrac{p-q+3}{2}\Big]$

\section{Example: Kerr-Newman-de Sitter solution}

In this section we will consider one explicit example of the exact solution
of Einstein equations. This is the Kerr-Newman-(anti-)-de Sitter solution
for a rotating black hole with the cosmological term $\Lambda $. The metric
has a form: 
\begin{equation}
ds^{2}=-\dfrac{\Delta _{r}}{\Xi ^{2}\varrho ^{2}}\Big(dt-a\sin ^{2}\theta
d\varphi \Big)^{2}+\dfrac{\varrho ^{2}}{\Delta _{r}}dr^{2}+\dfrac{\varrho
^{2}}{\Delta _{\theta }}d\theta ^{2}+\dfrac{\Delta _{\theta }\sin ^{2}{%
\theta }}{\Xi ^{2}\varrho ^{2}}\Big(adt-(r^{2}+a^{2})d\varphi \Big)^{2},
\label{kds}
\end{equation}%
\begin{eqnarray}
\varrho ^{2} &=&r^{2}+a^{2}\cos ^{2}\theta , \\
\Delta _{r} &=&(r^{2}+a^{2})(1-\frac{1}{3}\Lambda r^{2})-2mr+e^{2}, \\
\Delta _{\theta } &=&1+\frac{1}{3}\Lambda a^{2}\cos ^{2}\theta , \\
\Xi &=&1+\frac{1}{3}\Lambda a^{2}.
\end{eqnarray}

We are interested in multiple roots. It is easy to check that a quadruple
root is impossible independently of the sign of $\Lambda $. The triple root
is possible, provided $\Lambda >0$. Therefore, hereafter we consider only
this case, i.e. the Kerr-Newman-de Sitter metric.

For a triple root, $p=q=3$, the function $\Delta _{r}$ has the form 
\begin{equation}
\Delta _{r}=-\dfrac{\Lambda }{3}(r-b)^{3}(r+r_{0}).
\end{equation}

By comparing with (\ref{kds}), one finds 
\begin{equation}
b=\dfrac{1}{\sqrt{2\Lambda }}\sqrt{1-\dfrac{x}{3}},~~~~r_{0}=\dfrac{3}{\sqrt{%
2\Lambda }}\sqrt{1-\dfrac{x}{3}},
\end{equation}%
where $\Lambda a^{2}=x$.

For such a triple root we have near the horizon:

\begin{equation}
A=\dfrac{\Delta _{r}}{\delta ^{2}}=-\dfrac{8\Lambda ^{2}}{\sqrt{2\Lambda }}%
\sqrt{1-\dfrac{x}{3}}\dfrac{u^{3}}{3-x+6x\cos ^{2}{\theta }}-2\Lambda ^{2}%
\dfrac{6x\cos ^{2}{\theta }+7x-21}{(3-x+6x\cos ^{2}{\theta })^{2}}%
u^{4}+o(u^{4}),
\end{equation}

\begin{eqnarray}
g_{\phi } &=&\dfrac{(5+3x)^{2}(3+x\cos ^{2}{\theta })\sin ^{2}{\theta }}{%
2\Lambda (3+x)^{2}(3+2x+3x\cos {2\theta })}+ \\
&&+u\dfrac{6}{\sqrt{2\Lambda }}\sqrt{1-\dfrac{x}{3}}\dfrac{(3+5x)(3+x\cos
^{2}{\theta })(3-x+6x\cos {2\theta })}{(3+x)^{2}(3+2x+3x\cos {2\theta })^{2}}%
\sin ^{2}{\theta }+o(u).
\end{eqnarray}

This is consistent with expansion of general form (\ref{g2}) - (\ref{g3}).
Expansion for $\omega $ reads 
\begin{eqnarray}
\omega &=&\dfrac{6x}{a(3+5x)}-\dfrac{72x^{2}}{a^{3}\sqrt{2\Lambda }(3+5x)^{2}%
}\sqrt{1-\dfrac{x}{3}}u-\dfrac{324x^{2}(x-1)}{a^{3}(3+5x)^{3}}u^{2}+ \\
&&+\dfrac{144x^{3}}{a^{5}\sqrt{2\Lambda }}\sqrt{1-\dfrac{x}{3}}\dfrac{%
(36x^{2}\cos {2\theta }+31x^{2}+138x-45)}{(3+5x)^{4}(3+x\cos ^{2}{\theta })}%
u^{3}+o(u^{3}),
\end{eqnarray}

This expansion is consistent with expansion, obtained from (\ref{omega_exp})

For $N^{2}$ we have

\begin{eqnarray}
N^{2} &=&-\dfrac{72\Lambda ^{2}(3+2x+3x\cos {2\theta })}{\sqrt{2\Lambda }%
(3+x)^{2}(3+5x)^{2}}\sqrt{1-\dfrac{x}{3}}u^{3}- \\
&&-\dfrac{18\Lambda ^{2}(69x+2x^{2}-63+9x(7x-15)\cos {2\theta })}{%
(3+x)^{2}(3+5x)^{2}}u^{4}+o(u^{4}).
\end{eqnarray}

This is also consistent with our expansion (\ref{A_N^2_rel}). Note that 
\begin{equation}
A_{3}\kappa _{3}=96\dfrac{\Lambda ^{3}(3-x)}{(9+18x+5x^{2})^{2}}.
\end{equation}

This is constant, as our expansions (\ref{A_rel}) predict. Also for this
case the relations (\ref{A_rel}) between $A_{4}$ and $\kappa _{4}$, $A_{5}$
and $\kappa _{5}$ are satisfied (but we do not give them here, because they
are place-consuming).

\section{Summary and conclusions}

Thus we derived the regularity conditions for the axially symmetric rotating
black holes in a quite general form. Our consideration relies on the most
natural coordinates in which the behavior of the metric in the near-horizon
region are characterized by two integers $p$ and $q$ only. The corresponding
conditions are formulated in terms of the metric expansions near the
horizon. The requirement of regularity selects only some types of such
expansions.

In doing so, the conditions are derived from two groups of requirements: (i)
the finiteness of the Riemann curvature and other invariants, (ii) the
finiteness of separate components in the Riemann tenor in a free falling
frame. The second type of requirement is stronger, as is seen from Tables I
and II. Our analysis is carried out separately for different types of a
horizon - nonextremal, extremal, ultraextremal ones.

The relation between (i) and (ii) enables us to formulate, in principle, the
notion of naked horizon and introduce the condition of analyticity in the
manner similar to the spherically symmetric case. This is supposed to be
done elsewhere.

\appendix

\section{The behavior of the velocity near horizon}

In this section we discuss the behavior of the 4-velocity near horizon
relevant in our context. As the metric is invariant with respect to $t$ and $%
\varphi $ translations, corresponding conservation laws give us: 
\begin{equation}
u^{t}=\dfrac{X}{N^{2}},~\mathrm{where}~X=\mathcal{E}-\omega \mathcal{L}%
,~~~u^{\varphi }=\dfrac{\mathcal{L}}{g_{\varphi \varphi }}+\dfrac{\omega X}{%
N^{2}},
\end{equation}%
where $\mathcal{E}$ and $\mathcal{L}$ are the specific (per unit mass)
energy and the component of the angular momentum generated by rotation in $%
\phi $direction. Normalization for 4-velocity $u^{\mu }u_{\mu }=1$ entails: 
\begin{equation}
u^{r}=\sigma \sqrt{A}\dfrac{\sqrt{X^{2}-N^{2}(1+\mathcal{L}^{2}/g_{\varphi
}+g_{\theta }(u^{\theta })^{2})}}{N}.
\end{equation}

Here $\sigma $ is a sign showing direction of motion. Hereafter,we consider
only "usual" particles (without fine-tuning of parameters).

The component $u^{\theta }$ of the four-velocity can be defined from the
geodesics equation but for our analysis it will be sufficient to take a
natural assumption that $u^{\theta }$ is finite near horizon. This means
that $u^{r}\sim \dfrac{\sqrt{A}}{N}$ near horizon.

Now let us analyze behavior of a trajectory near the horizon in the OZAMO
frame. To do this, first of all, we have to compute the components of
3-velocity, defined by relation: 
\begin{equation}
V^{(i)}=\dfrac{h_{\mu }^{(i)}u^{\mu }}{h_{\mu }^{(0)}u^{\mu }}.
\end{equation}

Using (\ref{ozamo}-\ref{ozamo_fr_last}), we can get: 
\begin{equation}
V^{(1)}=\dfrac{\mathcal{L}N}{\sqrt{g_{\varphi }}X},~~~~~V^{(2)}=\dfrac{u^{r}%
}{X}\dfrac{N}{\sqrt{A}},~~~~~~V^{(3)}=\sqrt{g_{\theta }}\dfrac{u^{\theta }N}{%
X}.
\end{equation}

Angles in the $r\theta $ and $r\varphi $ planes are defined as 
\begin{equation}
\tan \psi =\dfrac{V^{(3)}}{V^{(2)}}\sim O(N),~~~~~~\tan \delta =\dfrac{%
V^{(1)}}{V^{(2)}}\sim O(N).
\end{equation}

So, both angles $\sim O(N).$

\section{The regularity of the components of the Riemann tensor near the
horizon}

\label{appendix_1}

In this appendix we list some combinations of the metric coefficients and
their derivatives whose near-horizon behavior follows from the regularity of
the corresponding Riemann tensor in the FZAMO frame. The sign of colon shows
which very component requires this behavior.

\begin{gather}  
\hat{R}_{0113}:\dfrac{\sqrt{A}}{N}\Big(\partial _{r}\gamma _{\theta
}\partial _{\theta }\gamma _{\varphi }-\dfrac{\partial _{\theta }A}{A}%
\partial _{r}\gamma _{\varphi }+\partial _{\theta }\gamma _{\varphi
}\partial _{r}\gamma _{\varphi }-g_{\varphi \varphi }\dfrac{\partial
_{\theta }\omega \partial _{r}\omega }{N^{2}}-2\dfrac{\partial _{r}\partial
_{\theta }g_{\varphi \varphi }}{g_{\varphi \varphi }}\Big)=O(1),~~
\label{comp_frs} \\
\hat{R}_{0101}:\Bigg(\Big(\dfrac{\partial _{\theta }A}{A}+\dfrac{\partial
_{\theta }N^{2}}{N^{2}}\Big)\partial _{\theta }\gamma _{\varphi }+g_{\varphi
\varphi }\Big(\dfrac{\partial _{\theta }\omega }{N}\Big)^{2}- \\
-Ag_{\theta \theta }\Big[\Big(\dfrac{\partial _{r}A}{A}-\dfrac{\partial_r N^2%
}{N^2}\Big)\partial _{r}\gamma _{\varphi }-(\partial _{r}\gamma _{\varphi
})^{2}+2\dfrac{\partial _{r}^{2}g_{\varphi \varphi }}{g_{\varphi \varphi }}%
\Big]\Bigg)- \\
-2\dfrac{\sqrt{A}\psi}{\sqrt{g_{\theta\theta}}}\Big(\partial_r\partial_{%
\theta}\gamma_{\varphi}+\partial_r\gamma_{\varphi}\partial_{\theta}\gamma_{%
\varphi}+\partial_r\gamma_{\varphi}\dfrac{\partial_{\theta} A}{A}%
+g_{\varphi\varphi}\dfrac{\partial_{\theta}\omega\partial_r\omega}{N^2}\Big)%
=O(N^{2}),~~ \\
\hat{R}_{0203}:\dfrac{\sqrt{A}}{N}\Big(\Big(\dfrac{\partial _{\theta }A}{A}-%
\dfrac{\partial _{\theta }N^{2}}{N^{2}}\Big)\partial _{r}N^2+2\partial
_{r}\partial _{\theta
}N^{2}-\partial_{\theta}N^2\partial_r\gamma_{\theta}-3g_{\varphi\varphi}%
\partial_r\omega\partial_{\theta}\omega\Big)=O(N^{2}),~~ \\
\hat{R}_{0303}:\Bigg(\Big(\dfrac{\partial _{\theta }A}{A}+\dfrac{\partial
_{\theta }N^{2}}{N^{2}}\Big)\partial _{\theta }\gamma _{\theta }+3\Big(%
\dfrac{\partial _{\theta }A}{A}\Big)^{2}-2\Big(\dfrac{\partial _{\theta
}^{2}A}{A}+\dfrac{\partial _{\theta }^{2}N^2}{N^2}\Big)+\Big(\dfrac{%
\partial_{\theta}N^2}{N^2}\Big)^2+ \\
+3g_{\varphi \varphi }\Big(\dfrac{\partial _{\theta }\omega }{N}\Big)%
^{2}+Ag_{\theta \theta }\Big(\Big(\dfrac{\partial _{r}A}{A}-\dfrac{\partial
_{r}N^{2}}{N^{2}}\Big)\partial _{r}\gamma _{\theta }-(\partial _{r}\gamma
_{\theta })^{2}+2\dfrac{\partial _{r}^{2}g_{\theta \theta }}{g_{\theta
\theta }}\Big)\Bigg)- \\
-2\sqrt{A g_{\theta\theta}}\psi\Big(\Big(\dfrac{\partial_{\theta}N^2}{N^2}-%
\dfrac{\partial_{\theta}A}{A}\Big)\dfrac{\partial_r N^2}{N^2}+\dfrac{%
\partial_{\theta}N^2}{N^2}\partial_r\gamma_{\theta}-2\dfrac{%
\partial_r\partial_{\theta}N^2}{N^2}+3g_{\varphi\varphi}\dfrac{%
\partial_r\omega\partial_{\theta}\omega}{N^2}\Big)=O(N^{2}),~~ \\
\hat{R}_{0313}:\partial _{\theta }(\gamma _{\theta }-3\gamma _{\varphi
})\partial _{\theta }\omega +\dfrac{\partial _{\theta }N^{2}}{N^{2}}\partial
_{\theta }\omega -2\partial _{\theta
}^{2}\omega-g_{\theta\theta}A\partial_{r}\gamma_{\theta}\partial_r\omega
=O(N^{2}),~~~  \label{R_0313_rel} \\
\hat{R}_{0103}:\dfrac{\sqrt{A}}{N}\Bigg(\partial _{\theta }\omega \partial
_{r}(\gamma _{\theta }-3\gamma _{\varphi })+2\partial _{\theta }\omega 
\dfrac{\partial _{r}N^{2}}{N^{2}}-\Big(\dfrac{\partial _{\theta }A}{A}+%
\dfrac{\partial _{\theta }N^{2}}{N^{2}}\Big)\partial _{r}\omega -2\partial
_{r}\partial _{\theta }\omega \Bigg)+ \\
-\dfrac{\psi}{4N\sqrt{g_{\theta\theta}}}\Bigg[\partial_{\theta}\omega\Big(%
\dfrac{\partial_{\theta}A}{A}-\dfrac{\partial_{\theta}N^2}{N^2}%
-\partial_{\theta}(\gamma_{\theta}-3\gamma_{\varphi})\Big)%
+2\partial_{\theta}^2\omega- \\
-A g_{\theta\theta}\Big(\partial_r\omega\Big(\dfrac{\partial_r A}{A}-\dfrac{%
\partial_r N^2}{N^2}-\partial_r(\gamma_{\theta}-3\gamma_{\varphi})\Big)%
+2\partial_r^2\omega\Big)\Bigg]=O(N^{2}).  \label{lst_conf}
\end{gather}

All other components give the conditions equivalent to those listed above.%

Also note that in the conditions listed above, the terms proportional to the
angle $\delta$ do not appear. The reason for this is clear from Appendix A
above. The fact that $\delta=O(N)$ makes regular all terms, proportional to $%
\delta$ in expressions for Riemann tensor, so corresponding terms do not
appear in (\ref{comp_frs}-\ref{lst_conf})

\section{Regularity conditions of metric coefficients}

In this Appendix we list the conditions which we have to impose on the
expansion coefficients to make the expression (\ref{A_N2_cond_1}) 
\begin{equation}
\Big(\dfrac{\partial _{\theta }A}{A}+\dfrac{\partial _{\theta }N^{2}}{N^{2}}%
\Big)=O(N^{2}),
\end{equation}%
self-consistent. This condition has to be held for $\hat{R}_{0303}$ and $%
\hat{R}_{0101}$ to be regular. Let us consider general expansions in a form 
\begin{gather}
A=A_{q}(\theta )u^{q}+A_{q+1}(\theta )u^{q+1}+o(u^{q+1}), \\
N^{2}=\kappa _{p}u^{p}+\kappa _{p+1}u^{p+1}+o(u^{p+1}).
\end{gather}

First of all note that this can be rewritten in a form: 
\begin{equation}
\Big(\dfrac{\partial _{\theta }A}{A}+\dfrac{\partial _{\theta }N^{2}}{N^{2}}%
\Big)=\partial _{\theta }(\ln A+\ln N^{2})=O(N^{2}).  \label{A_eq}
\end{equation}

Starting with the first term in the expansion of the left hand side, we have 
\begin{equation}
\dfrac{A_{q}^{\prime }}{A_{q}}+\dfrac{\kappa _{p}^{\prime }}{\kappa _{p}}%
=0\rightarrow A_{q}=C_{p}/\kappa _{p},~~~C_p=\mathrm{const},
\label{A_N2_l=0}
\end{equation}

where prime denotes derivative with respect to $\theta$. Note that the
equation for $\hat{R}_{0203}$ in (\ref{comp_frs}) leads to the same
expansion. To get equations for higher order coefficients, first of all we
write expansions for $\ln A$ 
\begin{eqnarray}
\ln A &=&\ln (A_{q}u^{q}+A_{q+1}u^{q+1}+...+A_{q+l}u^{q+l}+...)=, \\
&=&\ln \Big(A_{q}u^{q}\Big(1+\dfrac{A_{q+1}}{A_{q}}u+...+\dfrac{A_{q+l}}{%
A_{q}}u^{l}+...\Big)\Big)=, \\
&=&\ln A_{q}u^{q}+\ln \Big(1+\dfrac{A_{q+1}}{A_{q}}u+...+\dfrac{A_{q+l}}{%
A_{q}}u^{l}+...\Big).
\end{eqnarray}

Expanding this, we can write: 
\begin{equation}
\ln A=\ln A_{q}u^{q}+\dfrac{A_{q+1}}{A_{q}}u+\Big(\dfrac{A_{q+2}}{A_{q}}-%
\dfrac{1}{2}\Big(\dfrac{A_{q+1}}{A_{q}}\Big)^{2}\Big)u^{2}+o(u^{2}),
\end{equation}%
\begin{equation}
\partial _{\theta }\ln A=\dfrac{A_{q}^{\prime }}{A_{q}}+\Big(\dfrac{A_{q+1}}{%
A_{q}}\Big)^{\prime }u+\Big(\Big(\dfrac{A_{q+2}}{A_{q}}\Big)^{\prime }-%
\dfrac{1}{2}\partial _{\theta }\Big(\Big(\dfrac{A_{q+1}}{A_{q}}\Big)^{2}\Big)%
\Big)u^{2}+o(u^{2}).
\end{equation}

Similar expansion takes place for $N^{2}$: 
\begin{equation}
\partial _{\theta }\ln N^{2}=\dfrac{\kappa _{p}^{\prime }}{\kappa _{p}}+\Big(%
\dfrac{\kappa _{p+1}}{\kappa _{p}}\Big)^{\prime }u+\Big(\Big(\dfrac{\kappa
_{p+2}}{\kappa _{p}}\Big)^{\prime }-\dfrac{1}{2}\partial _{\theta }\Big(\Big(%
\dfrac{\kappa _{p+1}}{\kappa _{p}}\Big)^{2}\Big)\Big)u^{2}+o(u^{2}).
\end{equation}

The term with $u$ in equation (\ref{A_eq}) gives us 
\begin{equation}
\Big(\dfrac{A_{q+1}}{A_{q}}\Big)^{\prime }+\Big(\dfrac{\kappa _{p+1}}{\kappa
_{p}}\Big)^{\prime }=0\rightarrow \dfrac{A_{q+1}}{A_{q}}=-\dfrac{\kappa
_{p+1}}{\kappa _{p}}+C_{p+1},~~~C_{p+1}=\mathrm{const}\text{.}
\label{A_N2_l=1}
\end{equation}

The term with $u^{2}$ in equation (\ref{A_eq}) entails 
\begin{equation}
\dfrac{A_{q+2}}{A_{q}}+\dfrac{\kappa _{p+2}}{\kappa _{p}}=\dfrac{1}{2}\Big(%
\Big(\dfrac{\kappa _{p+1}}{\kappa _{p}}\Big)^{2}+\Big(\dfrac{A_{q+1}}{A_{q}}%
\Big)^{2}\Big)+C_{p+2},~~~C_{p+2}=\mathrm{const}.  \label{A_N2_l=2}
\end{equation}

To write a general expression for $A_{q+l}$, we have to obtain coefficient
at $u^{l}$ in the expansion of $\ln A$. To find this coefficient, we firstly
write expansion for the logarithm. Denoting $f=\sum_{l=1}^{\infty }\dfrac{%
A_{q+l}}{A_{q}}u^{l}$, we can write 
\begin{equation}
\ln (1+f)=\sum_{n=1}^{\infty }\dfrac{(-1)^{n+1}}{n}f^{n}.
\end{equation}

Using a general expression 
\begin{equation}
(x_{1}+x_{2}+...+x_{m}+...)^{n}=\sum_{k_{j}}\dfrac{n!}{%
k_{1}!k_{2}!...k_{m}!...}x_{1}^{k_{1}}x_{2}^{k_{2}}...x_{m}^{k_{m}}...,
\end{equation}%
where 
\begin{equation}
k_{1}+k_{2}+...+k_{m}+...=n,  \label{n_sum}
\end{equation}%
we can write: 
\begin{equation*}
f^{n}=\sum_{k_{j}}\dfrac{n!}{k_{1}!...k_{m}!...}\Big(\dfrac{A_{q+1}}{A_{q}}%
\Big)^{k_{1}}...\Big(\dfrac{A_{q+m}}{A_{q}}\Big)%
^{k_{m}}...u^{k_{1}+...+mk_{m}+...}.
\end{equation*}

Using this, we can explicitly write expansion of (\ref{A_eq}) 
\begin{gather}
\partial_{\theta}(\ln A+\ln N^2)=\dfrac{A_{q}^{\prime }}{A_{q}}+\dfrac{%
\kappa _{p}^{\prime }}{\kappa _{p}}+\Bigg(\Bigg(\dfrac{A_{q+1}}{A_q}\Bigg)%
^{\prime}+\Bigg(\dfrac{\kappa_{p+1}}{\kappa_p}\Bigg)^{\prime}\Bigg)u+...+ \\
+\Bigg(\sum_{n=1}^{l}\dfrac{(-1)^{n+1}}{n}\sum_{k_{j}}\dfrac{n!}{%
k_{1}!..k_{m}!..}\Bigg[\prod_{j=1}^{l}\Big(\dfrac{A_{q+j}}{A_{q}}\Big)%
^{k_{j}}+\prod_{j=1}^{l}\Big(\dfrac{\kappa _{p+j}}{\kappa _{p}}\Big)^{k_{j}}%
\Bigg]\Bigg)^{\prime}u^{l}+...=O(N^2).
\end{gather}

In each term summation is taken over such $\{k_{1},k_{2},...,k_{m},...\}$
that eq. (\ref{n_sum}) and 
\begin{equation}
\sum_{j=1}^{l}jk_{j}=l  \label{l_sum}
\end{equation}%
are to be satisfied. We require all coefficients at terms, proportional to $%
u^{l}$ with $l<p$ to vanish. Conducting this procedure and integrating over $%
\theta $, we get recurrent relation: 
\begin{equation}
\dfrac{A_{q+l}}{A_{q}}+\dfrac{\kappa _{p+l}}{\kappa _{p}}=\sum_{n=2}^{l}%
\dfrac{(-1)^{n}}{n}\sum_{k_{j}}\dfrac{n!}{k_{1}!..k_{m}!..}\Bigg[%
\prod_{j=1}^{l}\Big(\dfrac{A_{q+j}}{A_{q}}\Big)^{k_{j}}+\prod_{j=1}^{l}\Big(%
\dfrac{\kappa _{p+j}}{\kappa _{p}}\Big)^{k_{j}}\Bigg]+C_{p+l},  \label{A_rel}
\end{equation}%
where $\{C_{p},C_{p+1},...,C_{p+l},...C_{2p-1}\}$ are constants, $l=2,$ $3,$ 
$4$ ... The case $l=1$ is not included in this formula and is described by
eq. (\ref{A_N2_l=1}). The formula for $l=2$ agrees with (\ref{A_N2_l=2}).
The relation (\ref{A_rel}) works for all $2\leq l<p$.

Below, we give particular relations that may appear to be useful for
applications. The condition for $l=3$: 
\begin{equation}
\dfrac{A_{q+3}}{A_{q}}+\dfrac{\kappa _{p+3}}{\kappa _{p}}=A_{q+2}\dfrac{%
A_{q+1}}{A_{q}^{2}}+\kappa _{q+2}\dfrac{\kappa _{q+1}}{\kappa _{q}^{2}}-%
\dfrac{1}{3}\Big(\Big(\dfrac{A_{q+1}}{A_{q}}\Big)^{2}+\Big(\dfrac{\kappa
_{q+1}}{\kappa _{q}}\Big)^{2}\Big)+C_{p+3}.
\end{equation}%
The condition for $l=4$: 
\begin{eqnarray}
\dfrac{A_{q+4}}{A_{q}}+\dfrac{\kappa _{p+4}}{\kappa _{p}} &=&A_{q+3}\dfrac{%
A_{q+1}}{A_{q}^{2}}+\kappa _{p+3}\dfrac{\kappa _{p+1}}{\kappa _{p}^{2}}-\Big(%
A_{q+2}\dfrac{A_{q+1}^{2}}{A_{q}^{3}}+\kappa _{p+2}\dfrac{\kappa _{p+1}^{2}}{%
\kappa _{p}^{3}}\Big)+ \\
&&+\dfrac{1}{2}\Big(\Big(\dfrac{A_{q+2}}{A_{q}}\Big)^{2}+\Big(\dfrac{\kappa
_{p+2}}{\kappa _{p}}\Big)^{2}\Big)+\dfrac{1}{4}\Big(\Big(\dfrac{A_{q+1}}{%
A_{q}}\Big)^{4}+\Big(\dfrac{\kappa _{q+1}}{\kappa _{q}}\Big)^{4}\Big)+C_{p+4}
\end{eqnarray}

\end{document}